\pdfoutput=1
\documentclass[11pt]{article}

\usepackage{ACL2023}
\usepackage{enumerate}
\usepackage{times}
\usepackage{latexsym}
\usepackage{enumitem}
\usepackage{bm}
\usepackage[T1]{fontenc}
\usepackage[utf8]{inputenc}

\usepackage{microtype}

\usepackage{inconsolata}
\usepackage{microtype}
\usepackage{makecell}

\usepackage{graphicx} %use graph format
\usepackage{subfigure}
\usepackage{amsfonts}
\usepackage{amssymb} 
\usepackage{booktabs} %booktabs 宏包可以让我们创建没有竖线分隔的表格
\usepackage{amsmath} %这个宏包引入了一些改进的数学环境
\usepackage{hyperref} %超链接跳转
\usepackage{float}    % 使用float 宏包
\usepackage{array}
\usepackage{booktabs}
\usepackage{multirow}
\usepackage{longtable}
\usepackage{booktabs}%%加粗表格线
\usepackage{xcolor}%定义了一些颜色
\usepackage{colortbl,booktabs}%第二个包定义了几个*rule

\usepackage{url}
\usepackage{soul}

\newcommand{\tabincell}[2]{\begin{tabular}{@{}#1@{}}#2\end{tabular}}  %表格自动换行
\usepackage{url}
\title{Few-shot Joint Multimodal Aspect-Sentiment Analysis Based on Generative Multimodal Prompt}
\author{Xiaocui Yang$^{1,2}$, Shi Feng$^{1}$, Daling Wang$^{1}$, Sun Qi$^{2,3}$, Wenfang Wu$^{1}$, Yifei Zhang$^{1}$, \\
\textbf{Pengfei Hong}$^{2}$, \textbf{Soujanya Poria}$^{2}$ \\
  $^{1}$Northeastern University, $^{2}$Singapore University of Technology and Design, \\
  $^{3}$Nanjing University of Science and Technology \\ 
  \texttt{\{yangxiaocui, wenfang\}@stumail.neu.edu.cn},  \\
  \texttt{\{fengshi, wangdaling, zhangyifei\}@cse.neu.edu.cn}, \\
  \texttt{\{pengfei\_hong, sporia\}@sutd.edu.sg}, \texttt{319106003718@njust.edu.cn}
  }

\begin{document}
\maketitle
\begin{abstract}
We have witnessed the rapid proliferation of multimodal data on numerous social media platforms. Conventional studies typically require massive labeled data to train models for Multimodal Aspect-Based Sentiment Analysis (MABSA). However, collecting and annotating fine-grained multimodal data for MABSA is tough.
To alleviate the above issue, we perform three MABSA-related tasks with quite a small number of labeled multimodal samples. We first build diverse and comprehensive multimodal few-shot datasets according to the data distribution. To capture the specific prompt for each aspect term in a few-shot scenario, we propose a novel Generative Multimodal Prompt (GMP)\footnote{\url{
https://github.com/YangXiaocui1215/GMP}.} model for MABSA, which includes the Multimodal Encoder module and the N-Stream Decoders module. We further introduce a subtask to predict the number of aspect terms in each instance to construct the multimodal prompt.
Extensive experiments on two datasets demonstrate that our approach outperforms strong baselines on two MABSA-related tasks in the few-shot setting. 
\end{abstract}
\section{Introduction}
\begin{figure*}
\centering
\includegraphics[scale = 0.65]{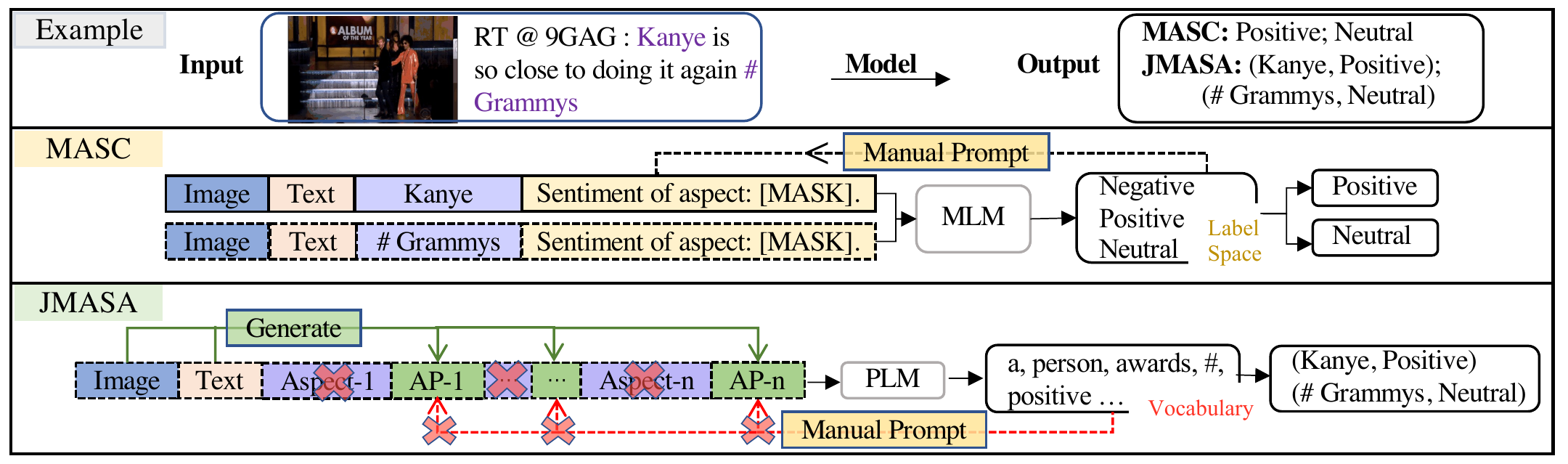} 
\vspace{-0.5em}
\caption{Examples of prompts for MASC and JMASA in a few-shot setting are shown. For MASC, the prompts consist of two triplets for the MABSA task, with aspect terms `Kanye' and `\#Grammy' shown in the upper part. The manual general prompt for the entire dataset can be efficient when the label space is limited, such as `Negative', `Positive', and `Neutral', as shown in the middle part with the yellow box representing the manual prompt for few-shot MASC in UP-MPF \cite{DBLP:conf/mm/YuZL22}. However, the manual prompt is infeasible for few-shot JMASA, as each aspect term is unknown and could be any phrase from the vocabulary, as shown by the red crosses. Hence, we generate a prompt for each aspect based on the multimodal context, as shown in the green box, where `AP' represents Aspect-oriented Prompt and `n' represents the number of aspect terms for an instance. `Vocabulary' refers to the vocabulary of the Pre-trained Language Model (PLM).
}
    \label{Fig_intro}
    \vspace{-1em}
\end{figure*}

The Multimodal Aspect-Based Sentiment Analysis (MABSA) task has garnered significant attention in recent times, as evidenced by several recent studies \cite{DBLP:journals/widm/ChandrasekaranN21, DBLP:journals/corr/abs-2203-01054, DBLP:journals/peerj-cs/ZhuXBXK22, DBLP:journals/inffus/GandhiAPCH23}. In the literature, MABSA is typically divided into three subtasks: Multimodal Aspect Term Extraction (MATE), Multimodal Aspect-oriented Sentiment Classification (MASC), and Joint Multimodal Aspect-Sentiment Analysis (JMASA) \cite{DBLP:conf/nlpcc/WuCWLC20, zhang2021multi, DBLP:conf/ijcai/Yu019, DBLP:conf/mm/0001F21, DBLP:conf/emnlp/JuZXLLZZ21, DBLP:conf/acl/LingYX22}. Given a text-image pair, MATE aims to extract all the aspect terms mentioned in the text, MASC focuses on detecting the sentiment corresponding to each extracted aspect term, and JMASA is designed to extract aspect terms and their corresponding sentiments jointly.
Previous studies on Multimodal Aspect-Based Sentiment Analysis (MABSA) primarily focus on leveraging extensive training data (full training datasets), with some works resorting to additional data to improve performance \cite{DBLP:conf/emnlp/JuZXLLZZ21, DBLP:conf/acl/LingYX22}. However, collecting and annotating such massive multimodal data for MABSA is time-intensive and laborious \cite{DBLP:journals/ijon/ZhouZHHH21}. Moreover, in real-world applications, only a limited amount of labeled data is commonly available. To address this challenge, PVLM \cite{DBLP:conf/icmcs/YuZ22} and UP-MPF \cite{DBLP:conf/mm/YuZL22} introduce prompt-based learning into Multimodal Aspect-oriented Sentiment Classification (MASC) in a few-shot scenario. Based on limited sentiment categories (three categories), PVLM and UP-MPF convert MASC to masked language modeling (MLM) tasks. However, the prerequisite of MASC is that the aspect terms are known, which requires aspect term extraction in advance, typically performed by Multimodal Aspect Term Extraction (MATE) or Joint Multimodal Aspect-Sentiment Analysis (JMASA). Both JMASA and MATE tasks are challenging due to the unknown and varying number of aspect items in each sample, as well as the distinct content of each aspect. Therefore, applying MLM in the few-shot setting is unsuitable for JMASA and MATE tasks, as depicted in Fig. \ref{Fig_intro}. This paper addresses the challenges of JMASA, MASC, and MATE in a text-image few-shot setting, and to the best of our knowledge, there are no dedicated studies dealing with JMASA and MATE tasks in the multimodal few-shot scenario.

Prior few-shot text classification tasks with limited classification labels have manually designed general prompts for the entire dataset to mine knowledge from pre-trained language models (PLM) \cite{DBLP:conf/emnlp/ShinRLWS20, DBLP:conf/naacl/Hosseini-AslLX22, DBLP:conf/naacl/ZhangFL022}. However, in the case of Joint Multimodal Aspect-Sentiment Analysis (JMASA) and Multimodal Aspect Term Extraction (MATE), where the content of each aspect term is unknown and assorted, manual prompts are infeasible for aspect extraction. To address this challenge, we propose a novel Generative Multimodal Prompt (GMP) model for few-shot Multimodal Aspect-Based Sentiment Analysis (MABSA), which includes the Multimodal Encoder (ME) module and the N-Stream Decoders (NSD) module. It is crucial to sample diverse and comprehensive data to build practical few-shot datasets in the multimodal few-shot setting.
We construct few-shot training and development datasets by sampling data with combinations of different sentiments in instances, according to the data distribution, as shown in Table \ref{Table_datasets}. Since the number of aspect terms in JMASA and MATE is unknown and vital, we leverage the Multimodal Encoder (ME) and Aspect-Num Decoder (AND) to predict the number of aspect terms as a subtask. The clues required for each aspect of an instance may vary. We generate aspect-oriented prompts for each aspect (aspect-level) using the ME and Aspect-oriented Prompt Decoder (APD). Similarly, we use the ME and Sentiment-oriented Prompt Decoder (SPD) to generate sentiment-oriented prompts. As the sentiment categories in all datasets are limited, we only reserve the instance-level sentiment prompts. The caption of the image modality is also captured as the image prompt. Lastly, specific multimodal prompts for different tasks are constructed based on the image caption, the predicted number of aspect terms, aspect prompts, and sentiment prompts. We feed the multimodal embedding with the multimodal prompt into the Multimodal Encoder-Decoder based BART model \cite{DBLP:conf/acl/LewisLGGMLSZ20} to generate triplet sequences.
Our main contributions are summarized as follows:
\vspace{-0.5em}
\begin{itemize} [leftmargin=*]
\item We propose a novel Generative Multimodal Prompt (GMP) model to handle Joint Multimodal Aspect-Sentiment Analysis (JMASA), Multimodal Aspect Sentiment Classification (MASC), and Multimodal Aspect Term Extraction (MATE) in the multimodal few-shot setting. To our knowledge, we are the first to focus on JMASA and MATE tasks in a multimodal few-shot scenario.
\item To tackle the challenge of unknown number of multimodal aspect terms and construct effective multimodal prompts, we employ multitasking and build the few-shot dataset by taking into account the distribution of sentiment categories for each dataset.
\item We conduct extensive experiments on the constructed few-shot datasets, and our results demonstrate that our proposed model outperforms strong baselines on JMASA and MASC in the few-shot setting.
\end{itemize}

% \hl{The main issue with this introduction is lack of motivation. Differences with the literature are not well explained. Please motivate the readers by presenting an illustrative figure that shows differences with existing models.}
% \vspace{-1em}
\section{Related Work}
\subsection{Multimodal Aspect Sentiment Analysis}
In contrast to coarse-grained sentiment analysis (sentence-level) \cite{DBLP:conf/acl/YangF0W20, DBLP:conf/naacl/LiXZZ22}, MABSA requires not only extracting aspect terms, but also recognizing the corresponding sentiment associated with each aspect. Early research has focused on different subtasks, including Multimodal Aspect Term Extraction (MATE) \cite{DBLP:conf/coling/SunWSWSZC20, DBLP:conf/acl/YuJYX20, DBLP:conf/mm/WuZCCL020, DBLP:conf/aaai/ZhangWLWZZ21, DBLP:journals/corr/abs-2205-03521} and Multimodal Aspect Sentiment Classification (MASC) \cite{DBLP:conf/iconference/YangYZN21, DBLP:conf/ijcai/Yu019, DBLP:conf/mm/0001F21}. More recently, Ju et al. \cite{DBLP:conf/emnlp/JuZXLLZZ21} proposed Joint Multimodal Aspect-Sentiment Analysis (JMASA), which jointly performs aspect term extraction and sentiment classification. Yang et al. \cite{DBLP:journals/ipm/YangNY22} introduced Cross-Modal Multitask Transformer (CMMT) for MABSA. VLP \cite{DBLP:conf/acl/LingYX22} further extends this by resorting to additional pre-training data and designing multiple pre-training tasks to enhance JMASA performance. However, few works have specifically addressed MABSA in the few-shot scenario. Although VLP has conducted low-resource experiments, it includes over 17,000 pre-training data and utilizes the full development dataset, which violates our starting point of adopting few-shot data.
\begin{figure*}[t] %%
  \centering %?????????
  \includegraphics[scale = 0.48]{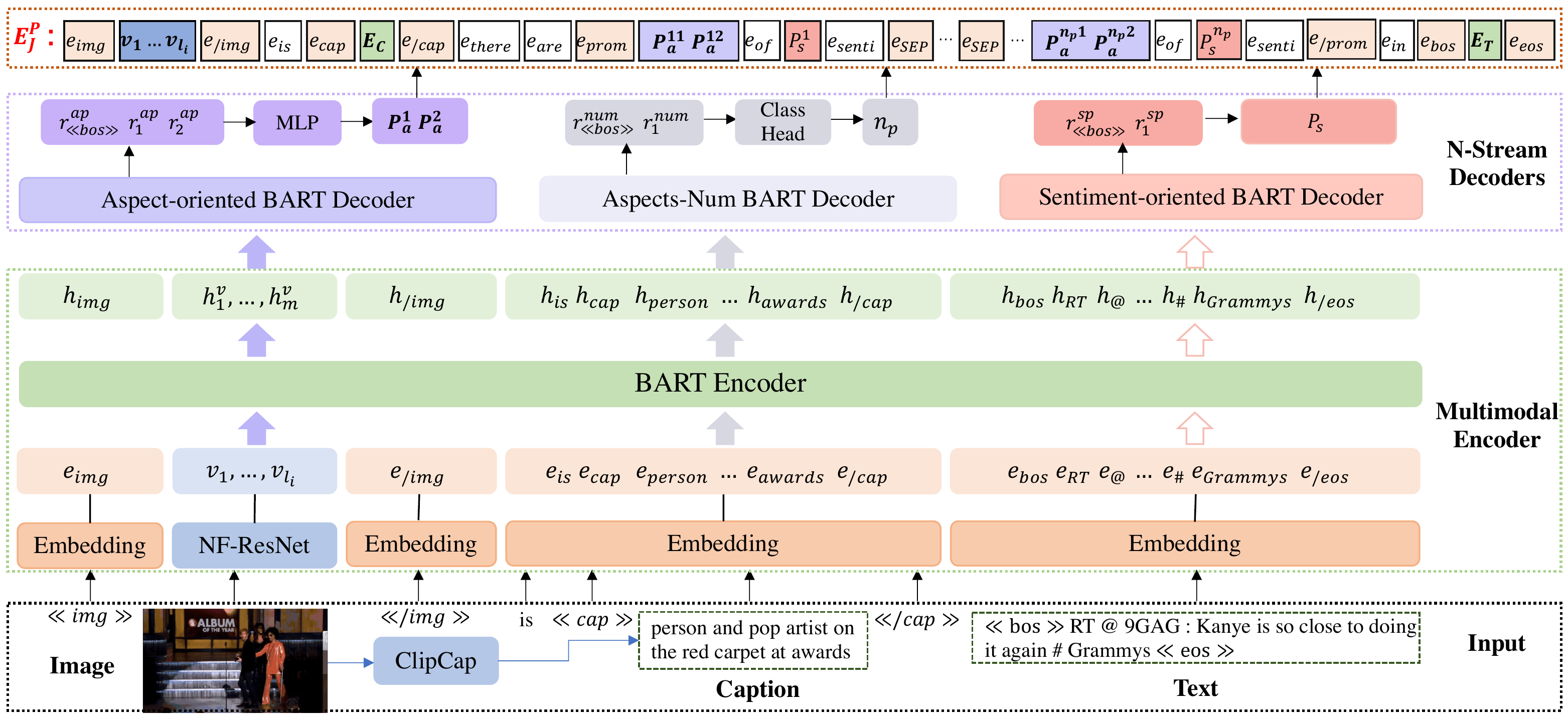} %????,??JPG,PNG,PDF,EPS?,???????
  \vspace{-0.5em}
\caption{The framework of our proposed Generative Multimodal Prompt (GMP) for Few-shot MABSA consists of two main modules: the Multimodal Encoder module (the green dashed box) and the N-Stream Decoders module (the purple dashed box).
For JMASA, we apply the multimodal embedding with the generative multimodal prompt $E_J^P$ using the Multimodal Encoder module. Similarly, for MASC and MATE, we design separate multimodal embeddings $E_S^P$ and $E_A^P$ respectively, as shown in Figure \ref{Fig_Prompt_for_MASC_and_MATE}.
The solid thick arrows (purple and gray) indicate the sharing of parameters between two multimodal encoders, while the thick hollow arrow (pink) does not share parameters with others. Special tokens such as $img$, $/img$, $cap$, $/cap$, $bos$, $eos$, $prom$, and $/prom$ are used in the multimodal embeddings. The embedding of ``sentiment" is denoted as $E_{senti}$.} 
\label{Fig_framwork} 
\vspace{-1em}
\end{figure*}
\vspace{-0.5em}
\subsection{Few-shot Learning with Pre-trained Language Model}
Prompt-based language modeling is applied to solve different few-shot tasks with PLM in Natural Language Process (NLP) due to its powerful representation \cite{DBLP:journals/corr/abs-2107-13586}, such as text classification \cite{DBLP:conf/emnlp/ShinRLWS20, DBLP:conf/naacl/Hosseini-AslLX22}, text regression \cite{DBLP:conf/acl/GaoFC20}, and text generation \cite{DBLP:conf/acl/LiL20}.
Existing works introduce Multimodal Prompt-based Fine-tuning (MPF) methods into multimodal settings by MLM, like Frozen \cite{DBLP:conf/nips/TsimpoukelliMCE21}, PVLM \cite{DBLP:conf/icmcs/YuZ22}, and UP-MPF \cite{DBLP:conf/mm/YuZL22}. 
Different from few-shot MASC (PLVM and UP-MPF), we simultaneously extract aspect terms and perform sentiment detection for each aspect in the multimodal few-shot scenario. 
\vspace{-0.5em}
\section{Our Proposed Model}
In Joint Multimodal Aspect-Sentiment Analysis (JMASA), our goal is to extract aspect terms and classify sentiment corresponding to each aspect. However, due to the varying number of aspect terms in each instance and each diverse aspect term, a different prompt is needed for each aspect in the few-shot setting. To address this, we propose a Generative Multimodal Prompt (GMP) for few-shot JMASA, as illustrated in Fig. \ref{Fig_framwork}. Leveraging BART, we generate aspect-oriented prompts for each aspect based on the multimodal context, as well as instance-level sentiment-oriented prompts.
\vspace{-0.5em}
\subsection{Task Formulation}
In this paper, we assume access to a pre-trained language model $\mathcal{M}$, such as BART, that we wish to fine-tune for the aspect-sentiment sequence generation task using labeled data. For the few-shot multimodal training dataset $\mathcal{D}_{train}$, we select $K$ training examples based on sentiment categories for each dataset, resulting in $\mathcal{D}{train} = {{(T^j, I^j, A^j, S^j, O^j)}}{j=1}^K$, where $T=[t^1, t^2, ..., t^{l_t}]$ is the text modality with $l_t$ as the text length; $I$ is the image modality; $A=[a^1, ..., a^n]$ is the aspect list; $S=[s^1, ..., s^n]$ is the sentiment list corresponding to $A$; and $O=[(x_b^1, x_e^1, s^1), ..., (x_b^n, x_e^n, s^n)]$ is our output, which represents the index-sentiment list, e.g., $O=[(5, 5, POS), (13, 14, NEU)]$ for the instance in Fig. \ref{Fig_2}. Here, $n$ denotes the number of aspects, $x_b^k$ and $x_e^k$ represent the beginning and end indices of the $k_{th}$ aspect term, and $s^k \in \{POS, NEG, NEU\}$ denotes the sentiment label.
For $\mathcal{D}_{dev}$, we select the same size of data as the few-shot training dataset, i.e., $|\mathcal{D}_{dev}| = |\mathcal{D}_{train}|$. Our task is to generate $O$ in the few-shot multimodal setting. Following the formulation in \cite{DBLP:conf/acl/YanDJQ020, DBLP:conf/acl/LingYX22}, we define the outputs of the three subtasks as follows\footnote{The underlined tokens are provided during inference.}:
\begin{itemize}
    \item \textbf{JMASA}:  $O=[(x^b_1, x^e_1, s_1), ..., (x^b_n, x^e_n, s_n)]$.
    \item \textbf{MASC}: $O=[(\underline{x^b_1}, \underline{x^e_1}, s_1), ..., (\underline{x^b_n}, \underline{x^e_n}, s_n)]$.
    \item \textbf{MATE}: $O=[(x^b_1, x^e_1), ..., (x^b_n, x^e_n)]$.
\end{itemize}
\vspace{-0.5em}
\subsection{Generative Multimodal Prompt}
GMP consists of two main modules: the Multimodal Encoder module and the N-Stream Decoders module.
% \vspace{-0.5em}
\subsubsection{\textbf{Multimodal Encoder}}
In this section, we design the multimodal encoder to capture multimodal representations. We start by extracting image representations using NF-ResNet \cite{DBLP:conf/iclr/BrockDS21}, and then project them to the text modality space for the image modality, $I$.
\begin{equation}
\begin{aligned}
     V &= Reshape(W_i ResNet(I) + b_i) \\
       &= [v^1, ..., v^k, ..., v^{l_i}], v^k \in \mathbb{R}^{d_t},
\label{equ_1}
\end{aligned}
\end{equation}
where $V$ is reshaped image representation, 
$W_i \in \mathbb{R}^{d_v \times d_{nt}}$, $b_i \in \mathbb{R}^{d_{nt}}$, and 
$nt = l_i\times d_t$. $l_i$, which is a hyperparameter, is the number of image slots that reserve initial image representation, and $d_t$ represents the dimension of text embedding in BART. 

Since BART is a pre-trained language model (PLM) that does not involve pre-training on image modality, we aim to alleviate the discrepancy issue of image representation in PLM. To achieve this, we further capture the image caption using ClipCap \cite{DBLP:journals/corr/abs-2111-09734}, denoted as $C$, which can be regarded as the image prompt.
\begin{equation}
     C = ClipCap(I).
    \label{equ_2}
\end{equation}
We utilize the BART model to obtain text embeddings for both the text input and the image caption.
\begin{equation}
\begin{aligned}
E_T = Embedding(T), E_T \in \mathbb{R}^{l_t \times d_t}, \\
E_C = Embedding(C), E_C \in \mathbb{R}^{l_{cap} \times d_t},
\label{equ_3}
\end{aligned}
\end{equation}
where $l_{cap}$ is the length of image caption.
The multimodal embedding $E_M$ can be obtained, 
$E_M$ $=$ $[E_{img}$, $V$, $E_{/img}$, $E_{is}$, $E_{cap}$, $E_C$, $E_{/cap}$, $E_{bos}$, $E_T$, $E_{eos}]$.
% \begin{equation}
% \begin{aligned}
% % E_M  = & [E_{\ll{img}\gg}, V, E_{\ll{/img}\gg}, E_{is}, E_{\ll{cap}\gg}, \\
% %     &E_C, E_{\ll{/cap}\gg}, E_{\ll{bos}\gg}, E_T, E_{\ll{eos}\gg}],
%     E_M  = & [E_{img}, V, E_{/img}, E_{is}, E_{cap}, E_C, E_{/cap}, \\
%           & E_{bos}, E_T, E_{eos}],
%     \label{equ_4}
% \end{aligned}
% \end{equation}
% where $img$, $/img$, $cap$, $/cap$, $bos$, and $eos$ are special tokens. 

Finally, we feed $E_M$ into the BART Encoder to obtain the multimodal representation. We argue that subsequent decoders require specific information, so we leverage different multimodal BART Encoders for this purpose.
\begin{equation}
    \begin{aligned}
        H^a_M = MBART^a_E(E_M), H^a_M \in \mathbb{R}^{l_m \times d},\\
        H^s_M = MBART^s_E(E_M), H^s_M \in \mathbb{R}^{l_m \times d}, 
        \label{equ_5}
    \end{aligned}
\end{equation}
% \begin{equation}
%     \begin{aligned}
%         H^s_M = MBART^s_E(E_M),
%         \label{equ_6}
%     \end{aligned}
% \end{equation}
where $l_m=l_i+l_{cap}+l_t+l_s$, $l_s$ is the length of special tokens, and $d$ is the hidden dimension. 
% \vspace{-0.5em}
\subsubsection{\textbf{N-Stream Decoders}}
In this section, we utilize the encoded multimodal representation from Eq. \ref{equ_5} to predict the number of aspect terms and generate aspect-oriented and sentiment-oriented prompts using different decoders for each instance. The 'N' in 'N-Stream' varies depending on the task, with values of 3, 2, and 1 for JMASA, MATE, and MASC, respectively.
% We construct the multimodal prompt based on the number of aspects and generated prompts.
\vspace{-0.5em}
\paragraph{Aspect-Num Decoder (AND).}
In the JMASA task, the number of aspects in each instance is significant but unknown, so we predict the number of aspects based on the multimodal context using the Aspect-Num BART Decoder as a subtask. Specifically, we input the multimodal encoder output $H^a_M$ and the special token $bos$ into the Aspect-Num Decoder, which then predicts the number of aspects $n_p\in \mathbb{R}^5$ as follows\footnote{Twitter 2017 dataset contains only 3 instances with more than 5 aspects. Therefore, we set "aspect-num" as 5 in the AND module to accommodate the maximum number of aspect terms in an instance.}:
\begin{equation}
    \begin{aligned}
        h^{and}_{n} = AND(H^a_M; E_{bos}),\\
        n_{p} = Softmax(MLP(h^{and}_{n})).
        \label{equ_7}
    \end{aligned}
\end{equation}
% \begin{equation}
%     \begin{aligned}
%        n_{p} = Softmax(MLP(h^{and}_{n})), n_p \in \mathcal{R}^5.
%         \label{equ_8}
%     \end{aligned}
% \end{equation}
We leverage the cross-entropy loss for the subtask,
\begin{equation}
    \begin{aligned}
      \mathcal{L}_{c} = - \sum_{j=1}^{K}{n^j_{g} log(n^j_p)},
        \label{equ_9}
    \end{aligned}
\end{equation}
where $n^j_g$ represents the label for the number of aspect terms. It's worth noting that in the MASC task, the gold number of aspect terms is provided to the model, and thus, this subtask is not required for MASC.

\begin{equation}
    \begin{aligned}
       P_a^k= MLP^{k}([h^{apd}_1, h^{apd}_2]),
        \label{equ_11}
    \end{aligned}
\end{equation}
where $k$ is the $k_{th}$ group of aspect of an instance, $ P_a^k \in \mathbb{R}^{2 \times d}$. 
The generative aspect-oriented prompt $AP=[P_a^1, ..., P_a^{n_p}] \in \mathbb{R}^{2n_p \times d}$.

\begin{figure}[t] %%
  \centering %?????????
  \includegraphics[scale = 0.42]{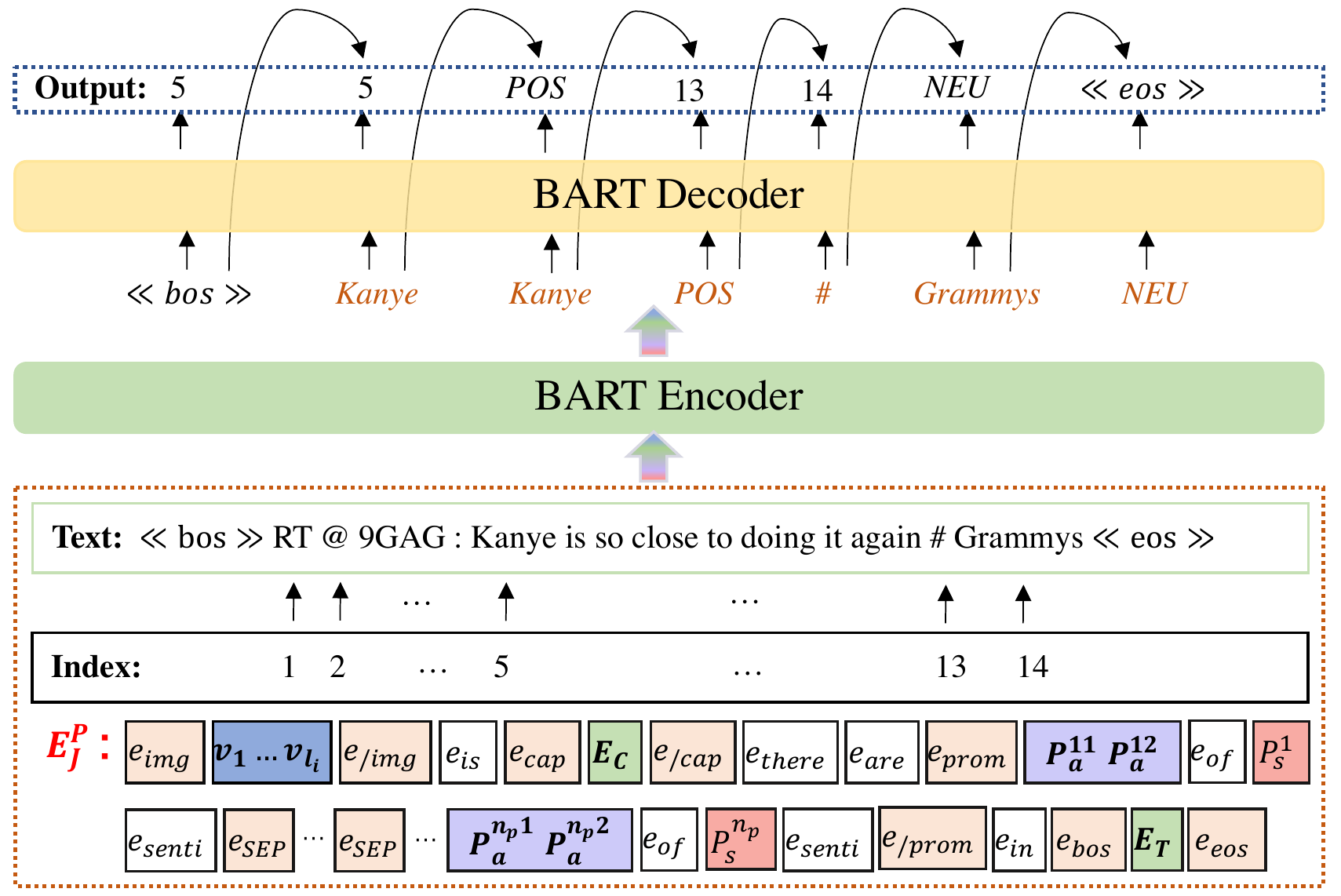} %????,??JPG,PNG,PDF,EPS?,???????
  \vspace{-0.5em}
\caption{An example of triplet sequence generation for the JMASA task.} 
\label{Fig_2} 
\vspace{-1em}
\end{figure}
\vspace{-0.5em}
\paragraph{Aspect-oriented Prompt Decoder (APD).}
Prompts for few-shot multimodal classification tasks can be manually designed for specific datasets due to limited categories, as demonstrated in PVLM \cite{DBLP:conf/icmcs/YuZ22} and UP-MPF \cite{DBLP:conf/mm/YuZL22}. However, each text-image pair carries different context information, and the aspects of the text are diverse. Therefore, in the few-shot setting, we need to capture various cues for each aspect. Inspired by this, we design our model to generate aspect-oriented prompts based on the multimodal context.
Specifically, we first generate an instance-level prompt based on the encoded multimodal representation. The final output of the JMASA task is a triplet sequence, where the first two positions of each triplet represent the beginning and ending indices for each aspect term. We set two aspect slots for each generated aspect-oriented prompt, resulting in an instance-level prompt length of $2n_p$. The decoder takes the encoder outputs $H^a_m$ and previous decoder outputs $h^{apd}{<{(l{ap}-1)}}$ as inputs to compute the current hidden state.
\begin{equation}
    \begin{aligned}
        h^{apd}_{l_{ap}} = APD(H^a_M; (h^{apd}_{<{(l_{ap}-1)}})),
        \label{equ_10}
    \end{aligned}
\end{equation}
where we feed the $bos$ into APD as the beginning token and $l_{ap}=2$.

% Each aspect in an instance is different, so we need to generate aspect-level prompt.
\vspace{-0.5em}
\paragraph{Sentiment-oriented Prompt Decoder (SPD).}
The sentiment corresponding to each aspect is related to each instance.
Similar to APD, we generate the sentiment-oriented prompt based on multimodal context.
For JMASA, the last position in each triplet of the output sequence predicts the sentiment.
We set one sentiment slot for each generated sentiment-oriented prompt, i.e., the length of the instance-level prompt is $n_p$.
\begin{equation}
    \begin{aligned}
        P_s =h^{spd}_{1} = SPD(H^s_M; E_{bos}),
        \label{equ_12}
    \end{aligned}
\end{equation}
where we feed the $bos$ into SPD as the beginning token. 
As the sentiment categories are limited, they share a common label space. Therefore, we do not generate corresponding sentiment cues for each aspect. Instead, $P_s$ is repeated $n_p$ times to form the generative sentiment-oriented prompt $SP=[P_s^1, ..., P_s^{n_p}] \in \mathbb{R}^{n_p \times d}$, where $d$ represents the dimensionality of the prompt.
\vspace{-0.5em}
\subsection{Multimodal Embedding with Prompt}
We construct the multimodal prompt for different tasks, including JMASA, MASC, and MATE, based on the text-image pair, aspect-oriented prompts, sentiment-oriented prompts, and prediction of the number of aspect terms.
For JMASA, we design multimodal embedding with a generative multimodal prompt, denoted as $\bm{E_J^P}$, as shown in Fig. \ref{Fig_framwork}.Similar to \bm{$E_J^P$}, we separately design multimodal embedding with prompt for MASC and MATE, e.g, \bm{$E_S^P$} and \bm{$E_A^P$} as Fig. \ref{Fig_Prompt_for_MASC_and_MATE} shows in Appendix \ref{Appendix_Multimodal_prompt}.
\subsection{Triplet Sequence Generation}
We next feed the multimodal embedding with prompt into the Encoder-Decoder model to generate the triplet sequence. We take the JMASA task as an example, as Fig. \ref{Fig_2} shows.
\begin{equation}
    \begin{aligned}
        H^P_J = MBART_E^{J}(E^P_J), H^P_J \in \mathbb{R}^{{l_J}\times d},
        \label{equ_16}
    \end{aligned}
\end{equation}
where $l_J$ is the length of $E^P_J$.
Then, we use the BART decoder to get the last hidden state,
\begin{equation}
    \begin{aligned}
        h^{dJ}_t = BART_D^{J}(H^P_J; \hat{O}_{<t}), 
        \label{equ_17}
    \end{aligned}
\end{equation}
where $t$ is the $t_{th}$ step and $\hat{O}_{<t}$ is the output of the previous $t$ steps.
Following \cite{DBLP:conf/acl/YanDJQ020}, we predict the token probability distribution $P_t$ with  $h_t^{dJ}\in \mathbb{R}^d$,  as follows:
\begin{equation}
    \begin{aligned}
        P_t = Predict([E_T; E_S]h^{dJ}_t), 
        \label{equ_18}
    \end{aligned}
\end{equation}
where $P_t \in \mathbb{R}^{l_t+l_c}$; $E_S$ is the embedding of the sentiment label set, and its length is $l_c=3$.

We employ cross-entropy loss for our sequence generation task.
\begin{equation}
    \begin{aligned}
      \mathcal{L}_{g} = - \sum_{j=1}^{K}{O^j log(P^j)}.
        \label{equ_19}
    \end{aligned}
\end{equation}
\vspace{-1em}
\subsection{Multitask Training}
We optimize our main task and subtask.
% The objective function is as follows:
\begin{equation}
    \begin{aligned}
      \mathcal{L} =  \mathcal{L}_g +  \lambda \mathcal{L}_c,
        \label{equ_20}
    \end{aligned}
\end{equation}
where $\lambda$ is the hyperparameter to control the contribution of each task.

\begin{table*}[t] \small
\begin{center}
\renewcommand{\arraystretch}{1} 
\begin{tabular}{
p{1.2cm}< \centering|
p{0.7cm}< \centering p{0.7cm}< \centering p{0.7cm}< \centering
p{1.7cm}< \centering p{1.8cm}< \centering p{1.7cm}< \centering p{2.5cm}< \centering p{1cm} < \centering }
\toprule[1pt]
\textbf{Datasets} & \textbf{POS} & \textbf{NEU} & \textbf{NEG} & \textbf{\{POS, NEU\}} & \textbf{\{NEG, NEU\}} & \textbf{\{POS, NEG\}} & \textbf{\{POS, NEU, NEG\}} & \textbf{All} \\
\midrule[1pt]
\textbf{15-Train} & \textbf{32}/526 & \textbf{64}/1084 & \textbf{16}/214 & \textbf{16}/178 & \textbf{8}/79 & \textbf{2}/13 & \textbf{0}/7  &
 \textbf{138}/2,101 \\
\textbf{15-Dev }& \textbf{32}/162 & \textbf{64}/375 & \textbf{16}/71 & \textbf{16}/69 & \textbf{8}/44 & \textbf{2}/6 & \textbf{0}/0 &
 \textbf{138}/727\\
\textbf{15-Test} & 167 & 335 & 68 & 73 & 28 &2 &1 & 674 \\
\midrule[1pt]
\textbf{17-Train }& \textbf{32}/534 & \textbf{32}/328 & \textbf{16}/150 & \textbf{32}/535 & \textbf{16}/153 & \textbf{2}/26 & \textbf{2}/20 &
 \textbf{132}/1,746 \\
\textbf{17-Dev} & \textbf{32}/177 & \textbf{32}/109 & \textbf{16}/49 & \textbf{32}/180 & \textbf{16}/50 & \textbf{2}/7 & \textbf{2}/5 & \textbf{132}/577\\
\textbf{17-Test} & 178 & 107 & 39 & 171 & 70 & 8 &14 & 587 \\
\bottomrule[1pt]
\end{tabular}
\vspace{-0.5em}
\caption{Statistics on two datasets. POS: Positive, NEU: Neutral, NEG: Negative. For A/B, B represents the number of original data, and A represents the number of few-shot data. \{Senti-1, Senti-2\} means that both Senti-1 and Senti-2 simultaneously exist in the instance, and there can be more than one of each sentiment.
For both datasets, the percentage of the constructed few-shot dataset accounts for about 7\% of the overall training data.
}
\label{Table_datasets}
\end{center}
\vspace{-1em}
\end{table*}
\vspace{-0.5em}
\section{Experiments}
We conduct experiments on two groups of few-shot multimodal datasets built according to the distribution of sentiment categories from Twitter-15 (15) and Twitter-17 (17) \cite{DBLP:conf/aaai/0001FLH18, DBLP:conf/acl/JiZCLN18}. We compare our model with numerous approaches on three tasks, including Multimodal Aspect Term Extraction (MATE), Multimodal Aspect-oriented Sentiment Classification (MASC), and Joint Multimodal Aspect-Sentiment Analysis (JMASA).
\vspace{-0.5em}
\subsection{Few-shot Datasets}
% We devote to the few-shot MABSA, so few-shot datasets need to be created.
To construct few-shot datasets for few-shot Multimodal Aspect-Based Sentiment Analysis (MABSA), it is important to select a few diverse samples that provide comprehensive coverage of the different sentiment categories. We sample data based on the distribution of sentiment categories in instances to create few-shot datasets. The statistics of the different datasets are presented in Table \ref{Table_datasets}. For each dataset, we randomly sample three groups of few-shot training and development datasets based on three different seeds, such as [42, 87, 100], and each split is run 3 times. We report the average performance and standard deviation over 9 (3 $\times$ 3) times of training for a more robust evaluation.
 \vspace{-0.5em}
\subsection{Implementation Details}
We utilize BART-Base with 140M parameters as our Pretrained Language Model (PLM), denoted as $\mathcal{M}$, and NF-ResNet-50 as our visual encoder. The number of epochs is set to 70, and the batch size is set to 4 for all tasks. The learning rates (lr) are set to 6.5e-5 for JMASA and MATE tasks, and for the MASC task, we set lr to 8e-5 and 7.5e-5 for Twitter-15 and Twitter-17, respectively. All models are implemented using PyTorch and the experiments are run on an A6000 GPU. Following \cite{DBLP:conf/acl/LingYX22}, we evaluate our model on three subtasks of MABSA and use Micro-F1 score (F1), Precision (P), and Recall (R) as the evaluation metrics to measure the performance. For MASC, we also use Accuracy (Acc) to compare fairly with other approaches. GMP has 169.3M/155.6M/154.9M parameters for JMASA/MATE/MASC, respectively, and during training, all parameters are updated. The training time for GMP up to 70 epochs is 50/50/25 minutes for JMASA/MATE/MASC.
\begin{table*}[t] \small
\begin{center}
\renewcommand{\arraystretch}{1} 
\begin{tabular}{
p{1.1cm}< \centering|
p{1.3cm}< \centering|
p{1.6cm}< \centering p{1.6cm}< \centering p{1.6cm}< \centering|
p{1.6cm}< \centering p{1.6cm}< \centering p{1.6cm}< \centering}
\toprule[1pt]
\multirow{2}{*}{\textbf{Modality}} & \multirow{2}{*}{\textbf{Model}} & 
\multicolumn{3}{p{5.6cm}<\centering|}{\textbf{Twitter-15}} & \multicolumn{3}{p{5.6cm}<\centering}{\textbf{Twitter-17}} \\
% \specialrule{0em}{2pt}{0pt} 
\cline{3-8}

% \cmidrule{3-8}
 % & & \makecell*[c]{P} & \makecell*[c]{\textbf{R}} &\makecell*[c]{\textbf{F1}} & \makecell*[c]{\textbf{P}} & \makecell*[c]{\textbf{R}} & \makecell*[c]{\textbf{F1}} \\
& & \textbf{P} & \textbf{R} & \textbf{F1} & \textbf{P} & \textbf{R} & \textbf{F1} \\
\midrule[1pt]
\multirow{3}{*}{\textbf{Text}}
& BART & 47.03\,($\pm$2.00) &	41.90\,($\pm$3.80) & 44.28\,($\pm$2.91) & 48.59\,($\pm$1.90)	& 44.97\,($\pm$1.95) & 46.70\,($\pm$1.81) \\
% & BART-GP$^\heartsuit$  & 46.30\,($\pm$2.21) &	43.23\,($\pm$1.15) & 44.71\,($\pm$1.64) & 51.61\,($\pm$2.73)& 	48.99\,($\pm$1.42) &	50.25\,($\pm$1.97)  \\
& D-GCN & 42.02\,($\pm$2.71) & 40.07\,($\pm$2.03) & 40.95\,($\pm$2.18) & 45.66\,($\pm$1.09) &	45.81\,($\pm$1.41) & 44.89\,($\pm$1.58) \\
& SpanABSA & 48.52\,($\pm$0.84) & 39.80\,($\pm$2.19) & 43.71\,($\pm$1.60) & 51.67\,($\pm$1.53) & 48.44\,($\pm$0.75) & 49.98\,($\pm$0.67)\\
\midrule[1pt]
\multirow{5}{*}{\textbf{\tabincell{c}{Text \\ -Image}}}
% & MBART$^\heartsuit$  & 49.24\,($\pm$2.28) &	46.00\,($\pm$0.91) &	47.55\,($\pm$1.54) & 52.49\,($\pm$3.02) &	50.03\,($\pm$0.50) &	51.20\,($\pm$1.71) \\
& JML & 48.51\,($\pm$1.14) &	41.59\,($\pm$2.56) &	44.77\,($\pm$1.97) & 50.13\,($\pm$0.41) & 48.65\,($\pm$0.10) & 49.38\,($\pm$0.25) \\
& CMMT & 29.85\,($\pm$1.37) & 36.23\,($\pm$2.05)  & 32.65\,($\pm$0.07)  & 39.64\,($\pm$0.51)  & 47.83\,($\pm$2.14)  & 43.34\,($\pm$1.12)  \\
& NVLP &46.04\,($\pm$0.82) &	42.40\,($\pm$0.25) &	44.14\,($\pm$0.47) & 50.66\,($\pm$2.09) &	45.92\,($\pm$1.10) & 48.16\,($\pm$1.28)   \\
& VLP & 46.56\,($\pm$0.94) &	\textbf{49.08\,($\pm$1.64)} &	47.77\,($\pm$0.73) & 51.32\,($\pm$0.19) & 52.22\,($\pm$0.52)	& 51.76\,($\pm$0.21) \\
% & MMBT &  0. & 0. & 0. & 0. & 0. & 0. \\
% & MVAN & 0.7298 & \textbf{0.7298} & 0.7236 & \textbf{0.7230} & 0.6646 & 0.6339 \\
& \textbf{GMP}& \textbf{51.67\,($\pm$2.01)} &	47.19\,($\pm$1.46) &	\textbf{49.33\,($\pm$1.71)} & \textbf{54.28\,($\pm$1.08)} &	\textbf{53.31\,($\pm$1.71)} &	\textbf{53.79\,($\pm$1.31)} \\
\bottomrule[1pt]
\end{tabular}
\vspace{-0.5em}
\caption{Average results of different models in terms of \textbf{Precision} (P), \textbf{Recall} (R), and \textbf{F1} for MABSA.
% $\heartsuit$ represents derivative models of our model.
}
\label{Table_JMASA}
\end{center}
\vspace{-2em}
\end{table*}

\subsection{Baselines}
To ensure a comprehensive comparison, we thoroughly evaluate our model against various approaches across different tasks.
\vspace{-0.5em}
\paragraph{Models for Joint Multimodal Aspect-Sentiment Analysis (JMASA).}
We first apply text-based approaches to perform Joint Aspect-Sentiment Analysis (JASA) with the following models:
\textbf{BART} \cite{DBLP:conf/acl/YanDJQ020} adapts JASA to an Encoder-Decoder model.
\textbf{D-GCN }\cite{DBLP:conf/coling/ChenTS20} proposes directional graph convolutional networks for JASA.
\textbf{SpanABSA} \cite{DBLP:conf/acl/HuPHLL19} applies an extraction-then-classification framework using a span-based labeling scheme.
Next, we accomplish JMASA and MATE using multimodal approaches with the following models:
\textbf{JML} \cite{DBLP:conf/emnlp/JuZXLLZZ21} performs JMASA by introducing auxiliary cross-modal relation detection.
\textbf{CMMT} \cite{DBLP:journals/ipm/YangNY22} proposes a multi-task learning framework that leverages two unimodal auxiliary tasks.
\textbf{VLP} \cite{DBLP:conf/acl/LingYX22}, which designs multiple Vision-Language pre-training tasks, is the state-of-the-art (SOTA) model for JMASA. However, since VLP introduces additional 17,000+ pre-training data, which violates our motivation to use few-shot data, we also present results for \textbf{NVLP}, which does not perform the pre-training task.
\vspace{-0.5em}
\paragraph{Models for Multimodal Aspect Sentiment Classification (MASC).}
We reproduce multimodal approaches that are trained in full MSA datasets from published paper for MASC.
\textbf{TomBERT} \cite{DBLP:conf/ijcai/Yu019} models the intra-modality and inter-modality dynamics to improve the performance of MASC.
\textbf{CapTrBERT} \cite{DBLP:conf/mm/0001F21} constructs an auxiliary sentence, which is the translation of the image, to provide multimodal information to a language model. 
\textbf{KEF} \cite{DBLP:conf/coling/ZhaoWLDHC22} exploits adjective-noun pairs extracted from the image to improve the visual attention capability and sentiment prediction capability of the fine-grained MSA task
    % of the fine-grained MSA task.
\textbf{FITE} \cite{DBLP:conf/emnlp/YangZ022}, the state-of-the-art model for fine-grained MSA, leverages facial information from the image modality. 
% \vspace{-1em}

Additionally, we adapt and evaluate models originally designed for \textbf{few-shot text classification tasks} for multimodal aspect-based sentiment classification.
\textbf{LM-BFF} \cite{DBLP:conf/acl/GaoFC20} designs different text prompts based on each specific dataset and text demonstrations to solve few-shot text classification tasks.
\textbf{LM-SC} \cite{DBLP:conf/naacl/JianGV22a} further introduces supervised contrastive learning based on LM-BFF to few-shot text tasks. 
\textbf{GFSC} \cite{DBLP:conf/naacl/Hosseini-AslLX22} converts the classification task into a generation task and solves text classification tasks in the few-shot setting through a pre-trained generation model, namely GPT2 \cite{radford2018improving}. 
Recently, \textbf{a few multimodal sentiment classification models} in the few-shot setting have emerged.
\textbf{PVLM} \cite{DBLP:conf/icmcs/YuZ22} proposes a prompt-based vision-aware language modeling approach to MASC in a few-shot scenario.
\textbf{UP-MPF} \cite{DBLP:conf/mm/YuZL22} applies a unified pre-training for multimodal prompt-based fine-tuning model, which is the state-of-the-art model for few-shot MASC. 
\vspace{-0.5em}
\subsection{Experimental Results and Analysis}
\subsubsection{Results of JMASA}
Table \ref{Table_JMASA} presents the results of JMASA on few-shot multimodal datasets, and several key observations can be made.
We can make the following observations.
First, multimodal models generally outperform unimodal models. Among the multimodal models, JML and VLP, which leverage additional data for relation detection and pre-training, respectively, achieve better performance compared to NVLP, which does not involve pre-training tasks, indicating the effectiveness of pre-training tasks in improving model performance.
When considering the amount of data used by the models, it is more reasonable to compare our model with NVLP. Our model consistently outperforms NVLP across both datasets, indicating its superior performance. Notably, our model also outperforms the second-best model, VLP, by a significant margin, with 1.56 and 2.03 absolute percentage points in terms of F1 on Twitter-15 and Twitter-17, respectively.
The superior performance of our model can be attributed to several factors. First, the generative multimodal prompt, which is based on the multimodal context, enables the model to capture practical knowledge for each sample from the pre-trained language model. Second, the subtask information provides valuable clues for constructing the multimodal prompt, leading to improved performance in few-shot multimodal sentiment classification.
\vspace{-0.5em}
\subsubsection{Results of the MASC}
The results of the MASC task on few-shot multimodal datasets, in terms of accuracy (Acc), are presented in Table \ref{Table_MASC_ACC}, while the corresponding F1 results are shown in Table \ref{Table_MASC_F1} from Appendix \ref{Appendix_MASC_F1}. The models with ``$*$'' are specifically introduced for few-shot scenarios. Several key observations can be made from the results.
We can obtain the following observations.
In the multimodal few-shot setting, 
1) Our model demonstrates the best performance in the multimodal few-shot setting, indicating its superiority over other models in handling the challenges of limited labeled data.
2) Prompt-based methods outperform robust multimodal models, highlighting the effectiveness of prompt-based methods in low-resource scenarios. This suggests that leveraging prompt engineering techniques, such as our generative multimodal prompt, can lead to improved performance in few-shot MSA.
3) BART, which uses only the text modality, performs better than most multimodal models, indicating the strong performance of our base model. This suggests that the pre-trained language model, BART, provides a solid foundation for our multimodal model.

\begin{table}[t] \small
\begin{center}
\renewcommand{\arraystretch}{1} 
\begin{tabular}{
p{1.1cm}< \centering|
p{1.6cm}< \centering|
p{1.6cm}< \centering |
p{1.6cm}< \centering}
\toprule[1pt]
\textbf{Modality} & \textbf{Model} & \textbf{Twitter-15} & \textbf{Twitter-17} \\
\midrule[1pt]

\multirow{4}{*}{\textbf{Text}}
& BART  &	65.57\,($\pm$3.07) & 64.12\,($\pm$1.47) \\
& LM-BFF$^*$  & 64.87\,($\pm$0.40) & 52.08\,($\pm$0.54)   \\
& LM-SC$^*$  & 65.47\,($\pm$1.74) & 57.51\,($\pm$2.95)   \\
& GFSC$^*$  & 60.75\,($\pm$1.07)  & 61.72\,($\pm$0.16)   \\
% & BART-GP$^\heartsuit$   & 51.49\,($\pm$9.02) & 63.52\,($\pm$2.51) &	62.16\,($\pm$3.54) &  64.56\,($\pm$3.09)  \\
\midrule[1pt]
\multirow{10}{*}{\textbf{\tabincell{c}{Text \\ -Image}}}
% & MBART   & 0\,($\pm$0) & 0\,($\pm$0) & 0\,($\pm$0)	& 0\,($\pm$0) \\
& TomBERT & 61.78\,($\pm$3.27) &   59.97\,($\pm$2.30) \\
& CapTrBERT & 58.76\,($\pm$0.25) &	56.48\,($\pm$1.61) \\
& JML-SC  &  60.36\,($\pm$0.90)	& 61.62\,($\pm$0.45)\\
& CMMT-SC  & 43.75\,($\pm$2.90) & 51.94\,($\pm$2.11)    \\
& KEF  & 55.81\,($\pm$3.74) &  46.50\,($\pm$0.075)   \\
& FITE  & 63.11\,($\pm$0.53) & 60.89\,($\pm$1.40)    \\
& NVLP  & 63.84\,($\pm$1.49) & 62.72\,($\pm$2.95) \\
& VLP  & 59.34\,($\pm$1.35) & 60.24\,($\pm$1.61) \\
\specialrule{0em}{1pt}{1pt} 
\cline{2-4}
\specialrule{0em}{1pt}{1pt} 
% \midrule[1pt]
& PVLM$^*$	& 64.54\,($\pm$1.81) & 	61.45\,($\pm$2.31) \\
& UP-MPF$^*$ &	63.71\,($\pm$3.62) &	62.02\,($\pm$0.40)\\
& \textbf{GMP} & \textbf{67.06\,($\pm$0.55)} & \textbf{66.20\,($\pm$1.12)} \\
\bottomrule[1pt]
\end{tabular}
\vspace{-0.5em}
\caption{Results of different models in terms of \textbf{Acc} for MASC on two datasets. ``$*$'' means that the model is proposed for the few-shot task.
% $\heartsuit$ represents derivative models of our model.
}
\label{Table_MASC_ACC}
\end{center}
\vspace{-2em}
\end{table}
\vspace{-0.5em}
\subsubsection{Results of MATE}
Table \ref{Table_MATE} presents the results of the MATE task. Among the models, VLP achieves the best performance in MATE, although it deviates from our initial goal of applying low-resource data due to its reliance on additional data and multiple pre-training tasks on the MVSA-Multiple Dataset \cite{DBLP:conf/mmm/NiuZPE16}. Similarly, JML also leverages additional data to enhance its performance.
An interesting observation is that MASC performs poorly in VLP when compared to NVLP, despite VLP showing better performance on the MATE and JMASA tasks compared to NVLP. We hypothesize that the pre-training task of VLP may be more aligned with the MATE task, which in turn may have an impact on the performance of MASC.
\begin{table}[t] \small
\begin{center}
\renewcommand{\arraystretch}{1} 
\begin{tabular}{
p{1cm}< \centering|
p{2cm}< \centering|
p{1.6cm}< \centering |
p{1.5cm}< \centering}
\toprule[1pt]
\textbf{Modality} & \textbf{Model} & \textbf{Twitter-15} & \textbf{Twitter-17} \\
\midrule[1pt]

\textbf{Text} & BART & 66.67\,($\pm$3.17) &	70.12\,($\pm$1.73)  \\
\midrule[1pt]
\multirow{5}{*}{\textbf{\tabincell{c}{Text \\ -Image}}}
% & MBART$^\heartsuit$  & 49.24\,($\pm$2.28) &	46.00\,($\pm$0.91) &	47.55\,($\pm$1.54) & 52.49\,($\pm$3.02) &	50.03\,($\pm$0.50) &	51.20\,($\pm$1.71) \\
& JML-MATE & 71.95\,($\pm$4.30) & 82.14\,($\pm$1.20) \\
& CMMT-MATE  & 73.19\,($\pm$2.50) & 82.50\,($\pm$0.59)    \\
& NVLP-MATE & 65.95\,($\pm$1.83)  & 71.52\,($\pm$0.26) \\
& VLP-MATE  & \textbf{77.61\,($\pm$0.25)}  &  \textbf{83.35\,($\pm$0.53)}  \\
% & MMBT &  0. & 0. & 0. & 0. & 0. & 0. \\
% & MVAN & 0.7298 & \textbf{0.7298} & 0.7236 & \textbf{0.7230} & 0.6646 & 0.6339 \\
& \textbf{GMP}& 73.65\,($\pm$1.35) & 79.95\,($\pm$0.43) \\
\bottomrule[1pt]
\end{tabular}
\vspace{-0.5em}
\caption{Results of different models in terms of \textbf{F1} for MATE on two datasets. 
% $\heartsuit$ represents derivative models of our model.
}
\label{Table_MATE}
\end{center}
\vspace{-2em}
\end{table}
\vspace{-0.5em}
\subsection{Ablation Experiments}
We performed ablation experiments on the GMP model to assess the effectiveness of different modules. The results, as shown in Table \ref{Table_Ablation}, indicate that the complete GMP model consistently the best performance across all tasks.
First, we remove the image modality (w/o Image) and built generative prompts based only on the text modality. The model's performance in all tasks is adversely affected, indicating that the image modality is crucial for achieving high performance in few-shot MSA tasks.
Next, we only remove the image caption (w/o Caption) and retain the initial image features to evaluate the effectiveness of the image prompt. The results show that the image prompt contributes to the overall performance of the model, indicating its utility in capturing important information from the image modality.
We also conduct experiments where we remove the multitask module (w/o Multitask) and set the number of aspect terms to 5 for each instance in the JMASA and MATE tasks. The performance of the models is affected, indicating that the subtask-specific modules are effective in capturing aspect-related information and improving performance.
To verify the utility of the generative multimodal prompt, we remove the multimodal prompt (w/o Prompt) and use only the original text-image representation. The model's performance degraded, indicating that our proposed multimodal prompt is beneficial in providing valuable cues for the sentiment analysis task.
We further remove the generative aspect prompt (w/o GAP) to assess the importance of GAP. 
Interestingly, we observe that using generated sentiment prompts (GSP) resulted in better performance in the MASC task (w/o GSP), whereas we obtain the opposite result in the JMASA task (w/ GSP). This suggests that the generated aspect prompt provides sufficient information to the model, and GSP may introduce redundant information in the JMASA task. However, in the MASC task, GSP provides effective cues for sentiment classification.
We further experiment with different generated sentiment prompts (w DSPrompt) and find that the performance significantly decrease. There are two possible reasons for this observation.
First, the sentiment categories in our dataset are limited. When using generated sentiment prompts for each aspect, it may introduce noise and irrelevant information to MASC.
Second, the generated prompts for each aspect provide sufficient information to guide the model in capturing aspect-related sentiment information. 

\begin{table}[t] \small
\renewcommand{\arraystretch}{1} 
    \centering
    \begin{tabular}{
    p{0.9cm}< \centering| p{1.9cm}< \centering| 
    p{1.6cm}< \centering| p{1.6cm}< \centering}
         % \hline
      	 \toprule[1pt]
        \specialrule{0em}{1pt}{1pt}
        \textbf{Task} & \textbf{Model} & \textbf{Twitter-15} & \textbf{Twitter-17} \\
        \midrule[1pt]
        \multirow{7}{*}{\textbf{JMASA}}
        & w /  GSPrompt & 47.11\,($\pm$3.12)  & 52.43\,($\pm$1.35) \\
        &  w/o Multitask & 47.70\,($\pm$1.41) & 49.77\,($\pm$1.69) \\
        &  w/o Image & 44.71\,($\pm$1.64) & 50.25\,($\pm$1.97) \\
        &  w/o Caption & 47.31\,($\pm$1.12) & 52.11\,($\pm$1.16) \\
        &  w/o Prompt & 47.55\,($\pm$1.54) & 51.20\,($\pm$1.71) \\ 
        & w/o GAPrompt & 48.05\,($\pm$1.38) & 48.81\,($\pm$4.98)\\
         % w/o DAPrompt & 0\,($\pm$0) & \\
         & \textbf{GMP}  & \textbf{49.33\,($\pm$1.71)} &	\textbf{53.79\,($\pm$1.31)} \\
         \midrule
        \multirow{6}{*}{\textbf{MATE}}
         & w/o Multitask & 73.46\,($\pm$0.94) & 79.02\,($\pm$1.16)\\
         &  w/o Image & 68.54\,($\pm$0.99) & 74.41\,($\pm$3.19) \\
         &  w/o Caption &  72.06\,($\pm$1.52) & 78.91\,($\pm$1.49) \\
          & w/o Prompt & 72.55\,($\pm$0.93) & 79.01\,($\pm$0.90) \\
         &  w/o GAPrompt & 71.62\,($\pm$0.71) & 78.74\,($\pm$0.94)\\
         &  \textbf{GMP}  & \textbf{73.65\,($\pm$1.35)} & \textbf{79.95\,($\pm$0.43)} \\
         \midrule
          \multirow{6}{*}{\textbf{MASC}}
         & w/ DSPrompt & 64.48\,($\pm$3.47) & 64.17\,($\pm$1.31) \\
         &  w/o Image & 65.09\,($\pm$1.66) & 65.68\,($\pm$0.67) \\
         &  w/o Caption & 64.81\,($\pm$3.60) & 66.01\,($\pm$1.69) \\
         &  w/o Prompt & 62.75\,($\pm$1.18) & 64.34\,($\pm$1.76) \\ 
         &  w/o GSPrompt &  65.03\,($\pm$1.49) & 63.57\,($\pm$2.29)\\
         &  \textbf{GMP}  & \textbf{67.06\,($\pm$0.55)} & \textbf{66.20\,($\pm$1.12)} \\
        % \hline
        \bottomrule[1pt]
    \end{tabular}
    \vspace{-0.5em}
    \caption{Ablation experiment results on three tasks, including JMASA, MATE, and MASC. ``w/'' indicates ``with'' and ``w/o'' indicates ``without''.}
    \label{Table_Ablation}
    \vspace{-2em}
\end{table} 
\vspace{-0.5em}
\subsection{Hyperparameters Setting}
The hyperparameter experiments of JMASA are shown in Fig. \ref{Fig_3}. The hyperparameter experiments on other tasks are in Appendix \ref{Appendix_hyperparameter}.

\textbf{Hyperparameters $l_i$ and $\lambda$ on JMASA}.
In order to effectively utilize image information through NF-ResNet, we conduct experiments with different settings of the hyperparameter $l_i$ in Eq. \ref{equ_1}, and the results are shown in Fig. \ref{Fig3.sub.a}.
We observe that our GMP model achieves the best performance on both datasets when the number of image slots, $l_i$, is set to 4. When $l_i$ is smaller, the image information is not fully utilized, and the model's performance is compromised. On the other hand, retaining more image features by setting a larger value for $l_i$ results in redundant information being provided to the model, which also leads to decreased performance.
When $l_i$ was set to 0, GMP only utilized the image prompt, i.e., the image caption $C$, and discarded the initial image representation $V$. 
We also employ the hyperparameter $\lambda$ to balance the contribution of the subtask, as shown in Fig. \ref{Fig3.sub.b}. We find that the best value of $\lambda$ varied across different datasets, with 0.1 being the optimal value for Twitter-15 and 0.15 for Twitter-17. When $\lambda$ is set to a larger value, the model's performance dramatically drop. This is because a larger value of $\lambda$ biases the model towards the subtasks, and we need to strike a balance among all tasks to achieve optimal performance.

\vspace{-1em}
\begin{figure}[H] %%
  \centering %????????
\subfigure[Comparisons of $l_i$.]{
  \label{Fig3.sub.a}
  \includegraphics[scale = 0.35]{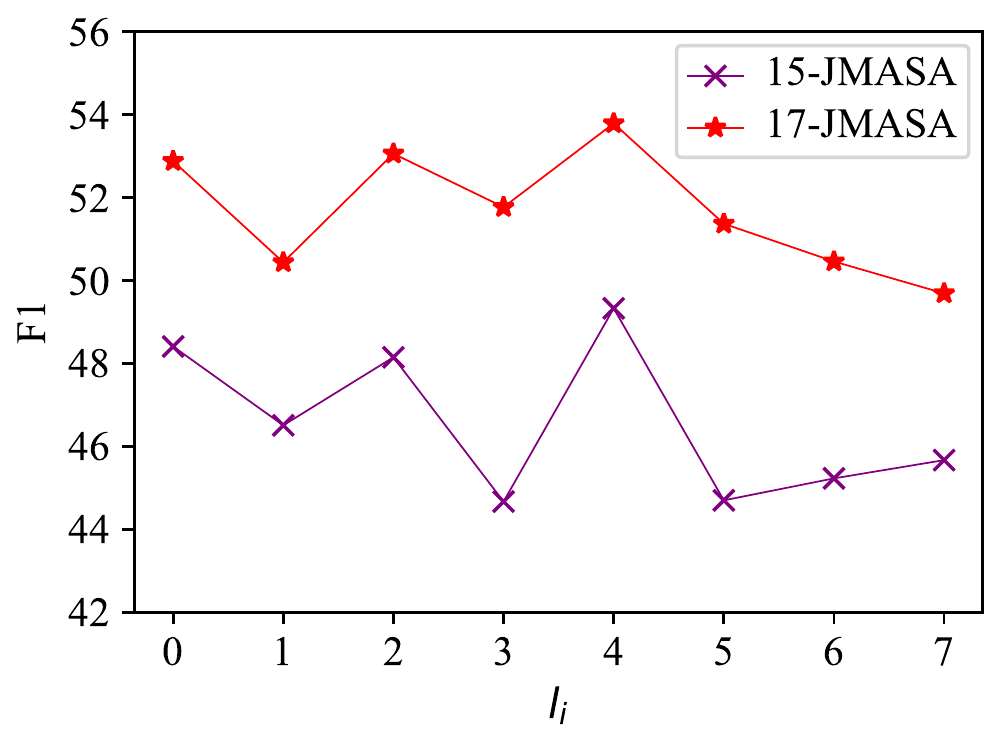}
}
\subfigure[Comparisons of $\lambda$.]{
  \label{Fig3.sub.b}
  \includegraphics[scale = 0.35]{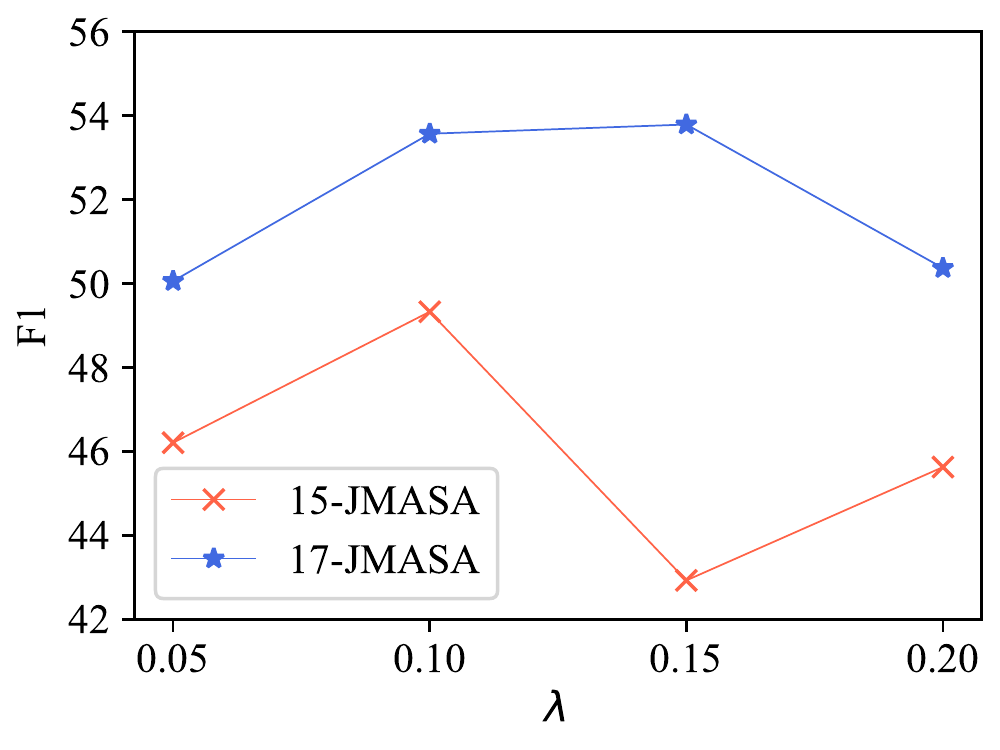}
}
\vspace{-1em}
\caption{\textbf{F1} comparisons of different Hyperparameters for JMASA.
} %?????
  \label{Fig_3} %
  \vspace{-1em}
\end{figure}

\vspace{-0.5em}
\section{Conclusion}
We propose a novel Generative Multimodal Prompt (GMP) for Multimodal Aspect-Based Sentiment Analysis (MABSA) that includes JMASA, MASC, and MATE in the multimodal few-shot scenario. We further introduce a subtask to predict the number of aspect terms to form multitask training to improve the performance of GMP.
Experimental results show that our proposed approach outperforms strong baselines on two subtasks of MABSA in the few-shot setting.
We provide a new direction for related tasks of MABSA in the few-shot setting.
In future work, we plan to exploit the fine-grained image features and achieve alignment between text and image modality to improve the performance of MABSA in the multimodal few-shot scenario. 
\section*{Limitations}
Although our model has shown superior performance, there are still a few limitations that could be improved in future work.
\begin{itemize}
\item We create few-shot datasets from the perspective of the combination of sentiment categories without considering the distribution of aspect items, such as the number of aspects in each sample. It may affect the performance of the model on the task of extracting aspects. We should create more efficient datasets for MABSA in the few-shot setting.
\item As we put more emphasis on the performance of the main task, the performance of the subtask of predicting the number of aspect terms in each example may suffer. We will further improve the accuracy of the subtask in future work.
\item We roughly exploit initial image features and do not perform alignment between text and image modalities. We plan to accomplish the alignment of multiple modalities further to improve the performance of MABSA in future work.
% \item In few-shot experiments, it is imperative to sample data. Selection of seeds may impact data sampling and model performance.
\end{itemize}

\section*{Acknowledgements}
Thanks to all co-authors for their hard work. The work is supported by National Natural Science Foundation of China (62172086, 62272092), Doctoral Research Innovation of Northeastern University (N2216004), and Chinese Scholarship Council.
\bibliography{acl2023}
\bibliographystyle{acl_natbib}

\appendix
\section{Multimodal Embedding with Prompt}
\label{Appendix_Multimodal_prompt}
\begin{figure*}[t] %%
  \centering %?????????
  \subfigure[The multimodal embedding with generative multimodal prompt for MASC.]{
  \includegraphics[scale = 0.55]{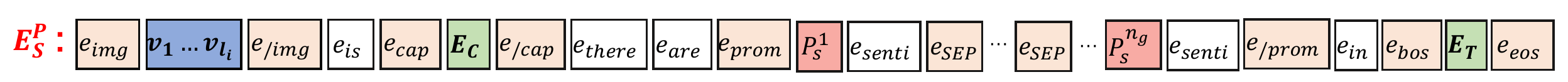} %????,??JPG,PNG,PDF,EPS?,???????
    \label{Fig_MASC_Prompt} 
    }
    \subfigure[The multimodal embedding with generative multimodal prompt for MATE.]{
     \includegraphics[scale = 0.55]{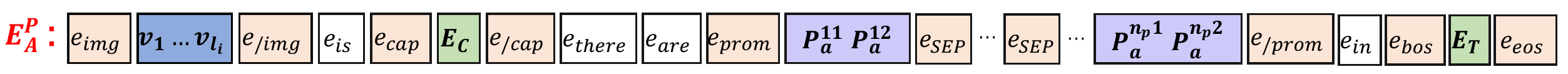} %????,??JPG,PNG,PDF,EPS?,???????
    \label{Fig_MATE_Prompt} }
    \caption{Multimodal embeddings with the generative multimodal prompt for MASC and MATE.}
    \label{Fig_Prompt_for_MASC_and_MATE}
\end{figure*}

For the MASC task, we design multimodal embedding with generative multimodal prompt, \bm{$E_S^P$}, as Fig. \ref{Fig_MASC_Prompt} shows.

For the MATE task, we design multimodal embedding with generative multimodal prompt, \bm{$E_A^P$}, as Fig. \ref{Fig_MATE_Prompt} shows.

\section{Experimental Results}
\subsection{F1 Results of MASC}
The results of the MASC task in terms of F1 are shown in Table \ref{Table_MASC_F1}. 
\label{Appendix_MASC_F1}
\subsection{Hyperparameters Setting}
\label{Appendix_hyperparameter}
\begin{figure}[h] %%
  \centering %????????
  \includegraphics[scale = 0.355]{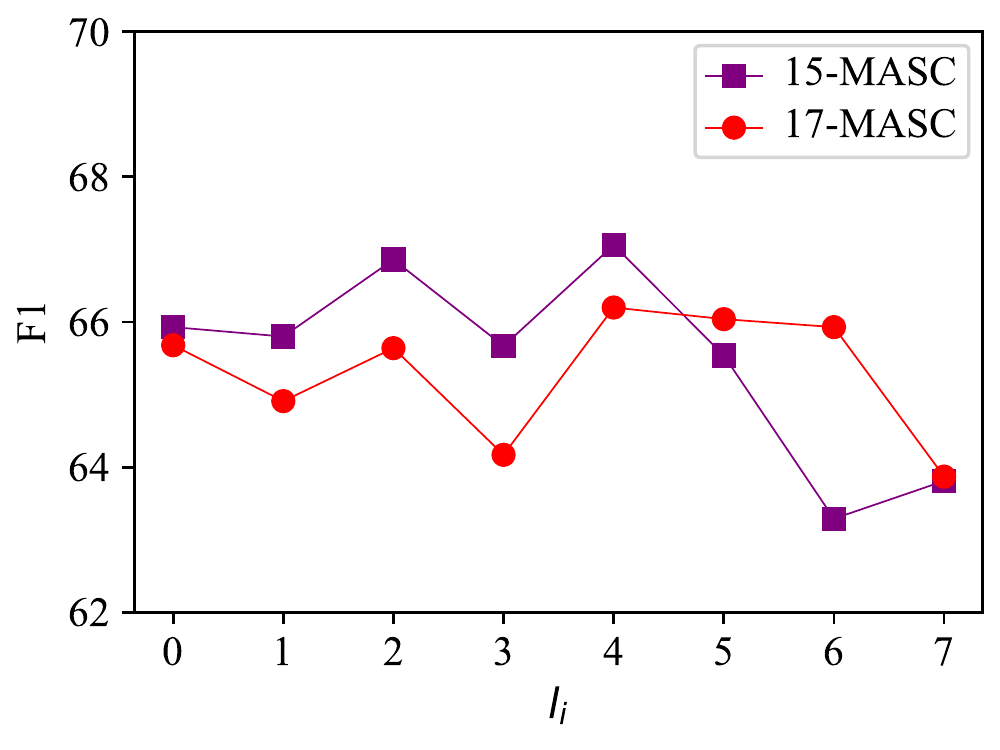}
  \vspace{-1em}
\caption{\textbf{Acc} comparisons of different Hyperparameters for MASC.
} %?????
  \label{Fig_4} %
  \vspace{-1em}
\end{figure}

\paragraph{Hyperparameters $l_i$ on MASC:}
We use the gold number of aspect terms for the MASC task and don't use the subtask. Thus we only conduct experiments on the hyperparameter $l_i$. Similar to the JMASA task, our model achieves the best performance on two datasets when $l_i$ is 4, as Fig. \ref{Fig_4} shows.

\begin{table}[H] \small
\begin{center}
\renewcommand{\arraystretch}{1} 
\begin{tabular}{
p{1cm}< \centering|
p{1.8cm}< \centering|
p{1.6cm}< \centering |
p{1.6cm}< \centering}
\toprule[1pt]
\textbf{Modality} & \textbf{Model} & \textbf{Twitter-15} & \textbf{Twitter-17} \\
\midrule[1pt]

\multirow{4}{*}{\textbf{Text}}
& BART  & 57.21\,($\pm$4.62) & 61.71\,($\pm$2.01)\\
& LM-BFF$^*$  & 58.27\,($\pm$1.46) & 49.04\,($\pm$3.40)   \\
& LM-SC$^*$  & 58.02\,($\pm$2.26) & 55.97\,($\pm$2.54)   \\
& GFSC$^*$   & 29.3\,($\pm$1.97) & 40.91\,($\pm$4.46)   \\
% & BART-GP$^\heartsuit$   & 51.49\,($\pm$9.02) & 63.52\,($\pm$2.51) &	62.16\,($\pm$3.54) &  64.56\,($\pm$3.09)  \\
\midrule[1pt]
\multirow{9}{*}{\textbf{\tabincell{c}{Text \\ -Image}}}
% & MBART   & 0\,($\pm$0) & 0\,($\pm$0) & 0\,($\pm$0)	& 0\,($\pm$0) \\
& TomBERT & 43.16\,($\pm$8.08)   &  54.92\,($\pm$2.40) \\
& CapTrBERT & 26.55\,($\pm$0.98) & 49.59\,($\pm$3.69) \\
& JML-SC  & 44.77\,($\pm$2.10) & 52.19\,($\pm$0.70)	\\
& CMMT-SC  & 45.52\,($\pm$0.85) & 51.92\,($\pm$1.00)    \\
& KEF  & 43.54\,($\pm$0.24) & 29.61\,($\pm$0.23)    \\
& FITE  & 58.97\,($\pm$0.34) & 59.16\,($\pm$2.15)    \\
& NVLP  & 55.11\,($\pm$2.20)  & 59.37\,($\pm$4.09)  \\
& VLP  & 44.56\,($\pm$3.83) & 56.09\,($\pm$2.43) \\
% \midrule[1pt]
\specialrule{0em}{1pt}{1pt} 
\cline{2-4}
\specialrule{0em}{1pt}{1pt} 
& PVLM$^*$  & 50.87\,($\pm$2.37)  & 59.62\,($\pm$1.81) \\
& UP-MPF$^*$  & 55.15\,($\pm$1.33) & 60.46\,($\pm$1.08)\\
% & MMBT &  0. & 0. & 0. & 0. & 0. & 0. \\
% & MVAN & 0.7298 & \textbf{0.7298} & 0.7236 & \textbf{0.7230} & 0.6646 & 0.6339 \\
& \textbf{GMP}& \textbf{60.31\,($\pm$1.83)} & \textbf{64.20\,($\pm$1.63)} \\
\bottomrule[1pt]
\end{tabular}
\vspace{-0.5em}
\caption{Results of different models in terms of \textbf{F1} for MASC on two datasets. 
% $\heartsuit$ represents derivative models of our model.
}
\label{Table_MASC_F1}
\end{center}
\end{table} 

\begin{figure}[H] %%
  \centering %????????
\subfigure[Comparisons of $l_i$.]{
  \label{Fig5.sub.a}
  \includegraphics[scale = 0.355]{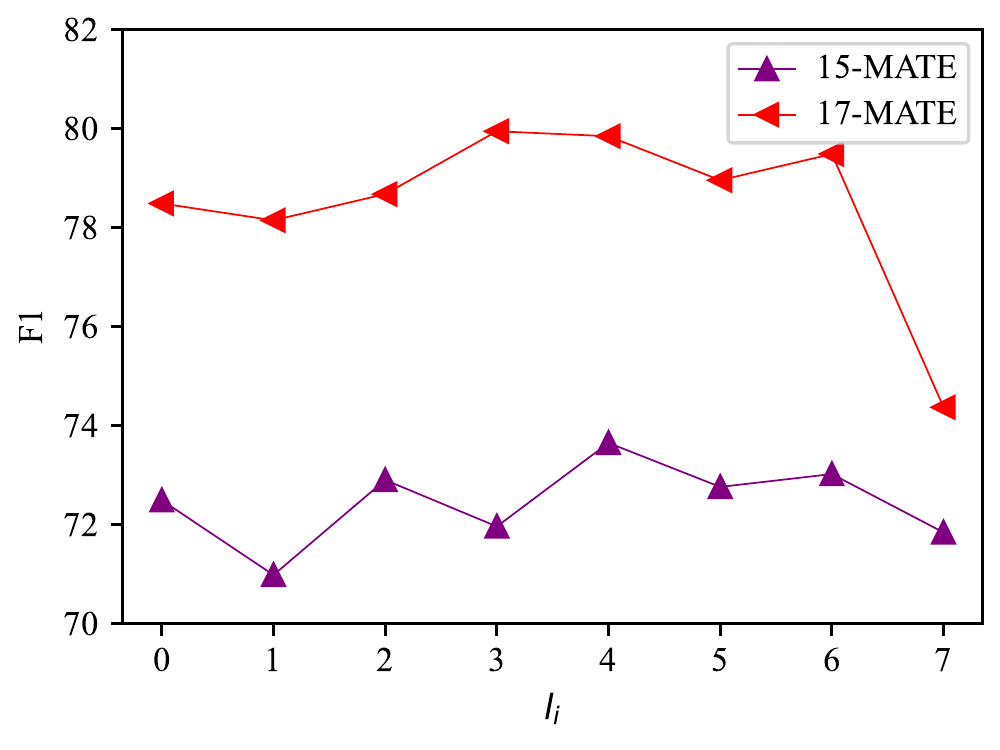}
}
\subfigure[Comparisons of $\lambda$.]{
  \label{Fig5.sub.b}
  \includegraphics[scale = 0.355]{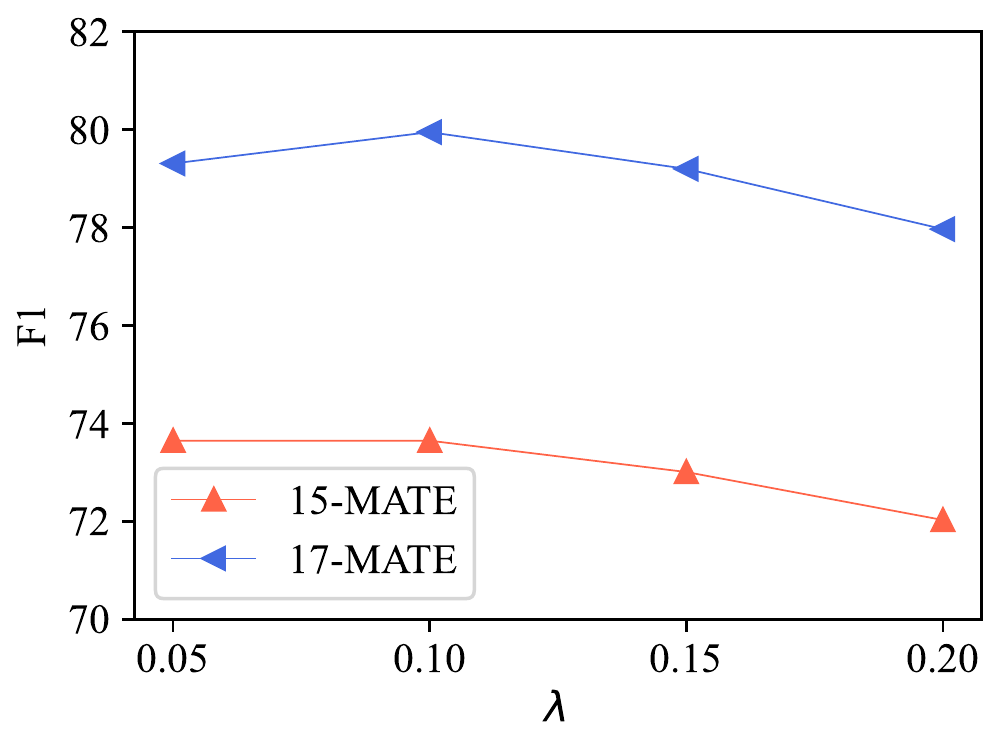}
}
\vspace{-1em}
\caption{\textbf{F1} comparisons of different Hyperparameters for MATE.
} %?????
  \label{Fig_5} %
\end{figure}

\paragraph{Hyperparameters $l_i$ and $\lambda$ on MATE:}
Fig. \ref{Fig_5} shows the hyperparameters of the MATE, including $l_i$ and $\lambda$.
On both datasets, our model has the best results when $\lambda$ is 4.
For the hyperparameter, $l_i$, our model achieves the best performance when $l_i$ is 4 on the Twitter-15 dataset, and $l_i$ is 3 on the Twitter-17 dataset.

\end{document}